\documentclass[aps,prb,twocolumn]{revtex4} 

\usepackage{graphicx}
\usepackage{dcolumn}
\usepackage{bm}

\newcommand{\comment}[1]{}

\begin{document}
\renewcommand{\theequation}{\arabic{section}.\arabic{equation}}

\title{Quantum Statistical Mechanics in Classical Phase Space.
Expressions for the Multi-Particle Density,
the Average Energy, and the Virial Pressure.}


\author{Phil Attard}
\affiliation{{phil.attard1@gmail.com}}


\begin{abstract}
Quantum statistical mechanics is formulated
as an integral over classical phase space.
Some details of the commutation function for averages are discussed,
as is the factorization of the symmetrization function
used for the grand potential and for the multi-particle density.
Three binary choices (eight routes) for the average energy
are shown to be mutually consistent.
An expression for the phase space function
that gives the average virial pressure is derived.
\end{abstract}


\maketitle

%
\section{Introduction}
\setcounter{equation}{0} \setcounter{subsubsection}{0}
%

For some years now I have been seeking to formulate
quantum statistical mechanics
as an integral over classical phase space.
\cite{QSM,Attard17,Attard16,STD2}
The motivation for this quixotic quest
is not just to pursue a different research direction,
but it comes also from the observed state of our physical world,
and from the computational challenges posed by quantum statistical mechanics.

In the terrestrial sphere most systems
are dominated by classical behavior.
For example, water at standard temperature and pressure
has a quantum correction on the order of one part in $10^4$,
depending upon the property of interest.
Hence it seems sensible to describe terrestrial condensed matter
using classical statistical mechanics
with added quantum corrections or perturbations.

From the computational viewpoint
classical statistical mechanics is easy,
whereas quantum statistical mechanics is difficult.
The chief technical barriers that inhibit
accurate numerical description
are the need for explicit eigenvalues and eigenfunctions,
and the need to either symmetrize the wave function
or else to enforce the occupancy rules for bosons and fermions.
In addition,
it is the rapid increase in computational cost with system size
that can be prohibitive.\cite{Bloch08,Hernando13}
It is routine nowadays for simulations of classical systems
to include tens of thousand of particles in three dimensions.
By comparison,
a typical exact simulation of a quantum system
comprises five particles in one dimension.\cite{Hernando13}
Of course, by introducing various approximations
one can treat larger systems.
For example, 6,500 Lennard-Jones atoms were able to be simulated
by using a variational approach,
approximating the eigenfunctions,
restricting the calculations to the ground state,
and neglecting wave function symmetrization.\cite{Georgescu11a}
In pursuing exact quantitative simulation results,
these examples suggest that a computationally efficient alternative
would be to formulate the problem in classical terms,
and to build in the quantum phenomena as systematic corrections
or perturbations.

This paper summarizes the current state
of the classical phase space formulation of quantum statistical mechanics
with a transparent derivation and precise notation.
In addition it resolves certain questions of consistency
in the expressions for the average energy.
Finally, an expression is derived for the phase function
that is to be used to calculate the average virial pressure.

%
\section{Partition Function,
Commutation Function, and Average}
\setcounter{equation}{0} \setcounter{subsubsection}{0}
%

\subsection{Grand Partition Function}

I consider $N$ spinless particles in $d$ dimensions
confined to a hypercube of volume $V = L^d$.
Where it is necessary to be explicit I shall use the position representation,
${\bf r} = \{{\bf r}_1, {\bf r}_2, \ldots , {\bf r}_N \}$,
with ${\bf r}_j = \{ r_{jx}, r_{jy}, \ldots, r_{jd} \}$.
The unsymmetrized position and momentum eigenfunctions are
respectively\cite{Messiah61}
\begin{equation}
|{\bf q}\rangle = \delta({\bf r}-{\bf q})
, \mbox{ and }
|{\bf p}\rangle
=
\frac{e^{-{\bf p}\cdot{\bf r}/i\hbar}}{V^{N/2} } .
\end{equation}
The position eigenvalues belong to the continuum.
The momentum eigenvalues are discrete
with spacing $\Delta_p = 2\pi\hbar/L$ per particle per dimension;
\cite{Messiah61}
almost immediately the continuum limit for this will be taken.
Each set of eigenfunctions is complete.

I shall now use these to transform the grand partition function
to classical phase space, after which I shall discuss in detail each
stage of the derivation.
The grand partition function is
\begin{eqnarray}
\Xi^\pm
& = &
\mbox{TR}' \left\{ e^{-\beta \hat{\cal H}}  \right\}
\nonumber \\ & = &
\sum_{N=0}^\infty \frac{z^N}{N!}
\sum_{\hat{\mathrm P}} (\pm 1)^p
\sum_{\bf p}
\langle \hat{\mathrm P} {\bf p} | e^{-\beta \hat{\cal H}} |  {\bf p} \rangle
 \nonumber \\ & = &
\sum_{N=0}^\infty \frac{z^N}{N!}
\sum_{\hat{\mathrm P}} (\pm 1)^p
\sum_{\bf p} \int  \mathrm{d}{\bf q} \;
\langle \hat{\mathrm P} {\bf p} |  {\bf q} \rangle \,
\langle {\bf q} | e^{-\beta \hat{\cal H}} | {\bf p} \rangle
\nonumber \\ & = &
\sum_{N=0}^\infty \frac{z^N}{h^{dN} N!}
\sum_{\hat{\mathrm P}} (\pm 1)^p
\! \int \!\! \mathrm{d}{\bf \Gamma} \,
\frac{ \langle {\bf q} | e^{-\beta \hat{\cal H}} | {\bf p} \rangle
}{ \langle {\bf q} |  {\bf p} \rangle }
\frac{ \langle \hat{\mathrm P} {\bf p} |  {\bf q} \rangle
}{ \langle {\bf p} |  {\bf q} \rangle }
\nonumber \\ & \equiv &
\sum_{N=0}^\infty \frac{z^N}{h^{dN} N!}
\int \mathrm{d}{\bf \Gamma}\;
e^{-\beta{\cal H}({\bf \Gamma})}
W_{p}({\bf \Gamma}) \eta^\pm_q({\bf \Gamma}) .
\end{eqnarray}
Here $z = e^{\beta \mu}$ is the fugacity,
$\beta = 1/k_\mathrm{B}T$ is the inverse temperature,
$h$ is Planck's constant,
and ${\bf \Gamma} = \{{\bf p},{\bf q}\}$
is a point in classical phase space.

The first equality here has the appearance of
von Neumann trace form for the partition function.
\cite{Messiah61,Merzbacher70,Pathria72}
It arises from the collapse of the total wave function
due to entanglement of the sub-system with the reservoir
upon energy (and particle) exchange.\cite{QSM,STD2}
The prime on the trace signifies that it is over unique states,
which is to say that each distinct state can only appear once,
or if more than once each occurrence must have appropriately reduced weight.
Not all workers avert to this requirement.

The second equality writes the trace as a sum over momentum states.
It also symmetrizes the eigenfunctions
by summing over all particle permutations $\hat{\mathrm P}$,
with the upper (plus) sign for bosons,
the lower (minus) sign for fermions,
according to the parity $p$ of the permutation.
The factor of $N!$ in the denominator
corrects for the double counting
of the states due to this symmetrization.\cite{QSM,STD2}
This formulation of particle statistics is formally exact,
and carries over to the continuum
where it would otherwise be impossible to enforce
particle state occupancy rules.

The third equality  inserts the completeness condition
$\int \mathrm{d}{\bf q}\; |{\bf q} \rangle \, \langle {\bf q} |
=\delta({\bf r}'- {\bf r}'')$.
Introducing this to the left of the Maxwell-Boltzmann operator as here
produces an asymmetry in position and momentum that is discussed shortly.

The fourth equality transforms to the momentum continuum.
The factor from the momentum volume element, $\Delta_p^{-dN}$,
combines with the factor of $V^{-N}
=\langle {\bf q} |  {\bf p} \rangle \, \langle {\bf p} |  {\bf q} \rangle $
that cancels the denominator introduced here
to give the prefactor $h^{-dN}$.
This is now an integral over classical phase space.

The fifth equality writes the phase space integral
in terms of  the commutation function $W_p$,
and symmetrization function $\eta_q^\pm$,
which will  be defined explicitly next.
If at the third equality the completeness condition
had been inserted to the right of the Maxwell-Boltzmann operator
one would have ended up with position and momentum swapped in these subscripts,
\begin{equation} \label{Eq:Wq-etap}
\Xi^\pm
=
\sum_{N=0}^\infty \frac{z^N}{h^{dN} N!}
\int \mathrm{d}{\bf \Gamma}\;
e^{-\beta{\cal H}({\bf \Gamma})}
W_{q}({\bf \Gamma}) \eta^\pm_p({\bf \Gamma}) .
\end{equation}
Obviously the partition function must remain unchanged.

The symmetrization functions are defined as
\begin{equation}
\eta^\pm_q({\bf p},{\bf q})
\equiv
\frac{1}{\langle {\bf p} | {\bf q} \rangle }
\sum_{\hat{\mathrm P}} (\pm 1)^p \,
\langle \hat{\mathrm P} {\bf p} | {\bf q} \rangle ,
\end{equation}
and
\begin{equation}
\eta^\pm_p({\bf p},{\bf q})
\equiv
\frac{1}{\langle {\bf q} | {\bf p} \rangle }
\sum_{\hat{\mathrm P}} (\pm 1)^p \,
\langle \hat{\mathrm P} {\bf q} | {\bf p} \rangle .
\end{equation}
Since $ \langle \hat{\mathrm P} {\bf p} | {\bf q} \rangle
= \langle {\bf p} |\hat{\mathrm P}  {\bf q} \rangle$,
it is evident that
$\eta^\pm_p({\bf p},{\bf q}) =
\eta^\pm_q({\bf p},{\bf q})^*$.
Also, since
$|{\bf p}\rangle \equiv
{e^{-{\bf p}\cdot{\bf r}/i \hbar}}/{V^{N/2}}$,
$\eta^\pm_p({\bf p},{\bf q})^* =
\eta^\pm_p(-{\bf p},{\bf q})$.
These  symmetrization functions will be discussed in greater detail
in \S \ref{Sec:loop}.

The commutation functions,
which are essentially the same as the functions
introduced by Wigner\cite{Wigner32}
and analyzed by Kirkwood,\cite{Kirkwood33}
are defined by
\begin{equation}\label{Eq:def-Wp}
e^{-\beta{\cal H}({\bf p},{\bf q})}
W_{p}({\bf p},{\bf q})
=
\frac{\langle{\bf q}|e^{-\beta\hat{\cal H}} |{\bf p}\rangle
}{\langle{\bf q}| {\bf p}\rangle } ,
\end{equation}
and
\begin{equation}
e^{-\beta{\cal H}({\bf p},{\bf q})}
W_{q}({\bf p},{\bf q})
=
\frac{\langle{\bf p}|e^{-\beta\hat{\cal H}} |{\bf q}\rangle
}{\langle{\bf p}| {\bf q}\rangle } .
\end{equation}
Again one has
$W_p({\bf p},{\bf q}) =
W_q({\bf p},{\bf q})^* $,
and also
$W_p({\bf p},{\bf q})^* =
W_p(-{\bf p},{\bf q})$.
High temperature expansions for the commutation function
have been given.\cite{Wigner32,Kirkwood33,STD2,Attard18}

The commutation function in phase space can also be written
as a series of energy eigenfunctions and eigenvalues,
$\hat{\cal H} |{\bf n}\rangle = E_n  |{\bf n}\rangle$.
Using the completeness properties of these
one obtains
\begin{eqnarray}
e^{-\beta{\cal H}({\bf p},{\bf q})}
W_{p}({\bf p},{\bf q})
& = &
\frac{\langle{\bf q}|e^{-\beta\hat{\cal H}} |{\bf p}\rangle
}{\langle{\bf q}| {\bf p}\rangle }
 \\ & = &
\frac{1}{\langle{\bf q}| {\bf p}\rangle }
\sum_{\bf n}
\langle{\bf q}|e^{-\beta\hat{\cal H}} |{\bf n}\rangle \,
\langle{\bf n} |{\bf p}\rangle
\nonumber\\ & = &
\frac{1}{\langle{\bf q}| {\bf p}\rangle }
\sum_{\bf n}
e^{-\beta E_{\bf n}} \,
\langle{\bf q}|{\bf n}\rangle \,
\langle{\bf n} |{\bf p}\rangle ,\nonumber
\end{eqnarray}
and analogously for $W_q$.
There is reason to believe
that this formulation in terms of energy (entropy)
eigenvalues and eigenfunctions
might be useful in computational applications.
\cite{Attard18}

It is clear that the grand partition function is real,
because the imaginary part of $W$ and of $\eta^\pm$ is odd in momentum.
Furthermore, as discussed for Eq.~(\ref{Eq:Wq-etap}),
the partition function is unchanged by the replacement
$W_p \eta^\pm_q \Rightarrow W_q \eta^\pm_p $.
Since these are the complex conjugate of each other,
this proves that the partition function is real.
On occasion below I shall omit the subscripts,
in which case it is implicit that whichever is used for $W$,
the opposite must be used for $\eta^\pm$.

In classical statistical mechanics
and in classical probability theory
is is essential that the weight density  be real and non-negative.
But in quantum statistical mechanics there is no such requirement.
Hence any problems associated with  $W$ and  $\eta^\pm$
being complex are practical and computational
rather than fundamental and principled.

\subsection{Statistical Averages}

The  grand canonical average of an operator
is similarly
\begin{eqnarray} \label{Eq:<A>}
\lefteqn{
\left< \hat A \right>_{z,\beta,V}^\pm
}  \\
& = &
\frac{1}{\Xi^\pm}
\mbox{TR}' \left\{ e^{-\beta \hat{\cal H}} \hat A \right\}
\nonumber \\ & = &
\frac{1}{\Xi^\pm} \sum_{N=0}^\infty \frac{z^N}{h^{dN} N!}
\int \mathrm{d}{\bf \Gamma}\;
e^{-\beta{\cal H}({\bf \Gamma})} A({\bf \Gamma})
W_{A,p}({\bf \Gamma}) \eta^\pm_q({\bf \Gamma})
\nonumber \\ & = &
\frac{1}{\Xi^\pm} \sum_{N=0}^\infty \frac{z^N}{h^{dN} N!}
\int \mathrm{d}{\bf \Gamma}\;
e^{-\beta{\cal H}({\bf \Gamma})} A({\bf \Gamma})
W_{A,q}({\bf \Gamma}) \eta^\pm_p({\bf \Gamma}). \nonumber
\end{eqnarray}
The final equality follows by swapping $ {\bf p}$ and $ {\bf q}$.
Reversing the order of
$e^{\beta \hat{\cal H}}$ and $\hat A$
replaces $W$ by $\tilde W$, here and in the defining equations next.
Such a cyclic permutation of the operators leaves
the trace and hence the average unchanged.

The commutation functions for the average are defined by
\begin{equation}
e^{-\beta{\cal H}({\bf p},{\bf q})} A({\bf p},{\bf q})
W_{A,p}({\bf p},{\bf q})
=
\frac{\langle{\bf q}|e^{-\beta\hat{\cal H}} \hat A |{\bf p}\rangle
}{\langle{\bf q}| {\bf p}\rangle } ,
\end{equation}
and
\begin{equation}
e^{-\beta{\cal H}({\bf p},{\bf q})} A({\bf p},{\bf q})
W_{A,q}({\bf p},{\bf q})
=
\frac{\langle{\bf p}|e^{-\beta\hat{\cal H}} \hat A |{\bf q}\rangle
}{\langle{\bf p}| {\bf q}\rangle } .
\end{equation}
Swapping the order of the operators gives $\tilde W$,
\begin{equation}
e^{-\beta{\cal H}({\bf p},{\bf q})} A({\bf p},{\bf q})
\tilde W_{A,p}({\bf p},{\bf q})
=
\frac{\langle{\bf q}|\hat A e^{-\beta\hat{\cal H}} |{\bf p}\rangle
}{\langle{\bf q}| {\bf p}\rangle } ,
\end{equation}
and
\begin{equation}
e^{-\beta{\cal H}({\bf p},{\bf q})} A({\bf p},{\bf q})
\tilde W_{A,q}({\bf p},{\bf q})
=
\frac{\langle{\bf p}|\hat A e^{-\beta\hat{\cal H}} |{\bf q}\rangle
}{\langle{\bf p}| {\bf q}\rangle } .
\end{equation}

In the simplest case the operator being averaged is an ordinary function
of the momentum and position operators,
$\hat A = A(\hat{\bf p},\hat{\bf q})$,
which leads directly to the phase space function $A({\bf p},{\bf q})$
appearing as here.
More generally, one should probably use the right hand side
to define the product $(A W_{A,p})({\bf p},{\bf q})$ etc.\
as a single phase space function.

I assume that the operator is Hermitian, $\hat A = \hat A^\dag$,
and so the average is real.
Taking the complex conjugate of the defining equation yields
\begin{eqnarray}
\lefteqn{
e^{-\beta{\cal H}({\bf p},{\bf q})} A({\bf p},{\bf q})
W_{A,p}({\bf p},{\bf q})^*
} \nonumber \\
& = &
\frac{\langle{\bf p}| \hat A  e^{-\beta\hat{\cal H}}|{\bf q}\rangle
}{\langle{\bf p}| {\bf q}\rangle }
\nonumber \\ & \equiv &
e^{-\beta{\cal H}({\bf p},{\bf q})} A({\bf p},{\bf q})
\tilde W_{A,q}({\bf p},{\bf q}).
\end{eqnarray}
Of course if $\hat A = A(\hat{\cal H})$,
it commutes with the Maxwell-Boltzmann operator.
In such a case
$\tilde W_{A,q} = W_{A,q}$, and  $\tilde W_{A,p} = W_{A,p}$.

In general
\begin{equation}
 W_{A,q}({\bf \Gamma}) \ne W_q({\bf \Gamma})
 \mbox{ and }
 W_{A,p}({\bf \Gamma}) \ne W_p({\bf \Gamma}).
\end{equation}
Unlike classical statistical mechanics,
for quantum  statistical mechanics the phase space weight
depends upon the function being averaged.
However in the case of pure functions,
from the definitions one can show that
\begin{eqnarray}
\mbox{ if }  \hat A = A(\hat{\bf p})
& \mbox{then}  &
W_{A,p}({\bf \Gamma})= W_{p}({\bf \Gamma})
\nonumber \\
& \mbox{and} &
\tilde W_{A,q}({\bf \Gamma}) = W_{q}({\bf \Gamma}).
\end{eqnarray}
And also that
\begin{eqnarray}
\mbox{ if }  \hat A = A(\hat{\bf q})
& \mbox{then}  &
W_{A,q}({\bf \Gamma}) = W_{q}({\bf \Gamma})
\nonumber \\
& \mbox{and} &
\tilde W_{A,p}({\bf \Gamma}) = W_{p}({\bf \Gamma}) .
\end{eqnarray}
One can see that any linear combination of pure functions
of the position and momentum operators
has an average that can be arranged
as a linear combination of averages with the original commutation function,
$ W_p $ or  $W_q $.
This is now illustrated explicitly for the energy operator.

\subsection{Energy} \label{Sec:<H>}

The first expression for the average energy takes
$\hat A = \hat{\cal H}$
and $W_{{\cal H},p}$ or  $W_{{\cal H},q}$,
giving
\begin{eqnarray} \label{Eq:<H>,WH}
\lefteqn{
\left< \hat {\cal H} \right>_{z,\beta,V}^\pm
}  \\
& = &
\frac{1}{\Xi^\pm}\sum_{N=0}^\infty \frac{z^N}{h^{dN} N!}
\int \mathrm{d}{\bf \Gamma}\;
e^{-\beta{\cal H}({\bf \Gamma})}
 {\cal H}({\bf \Gamma}) W_{{\cal H},p}({\bf \Gamma})
 \eta^\pm_q({\bf \Gamma})
\nonumber \\ & = &
\frac{1}{\Xi^\pm}\sum_{N=0}^\infty \frac{z^N}{h^{dN} N!}
\int \mathrm{d}{\bf \Gamma}\;
e^{-\beta{\cal H}({\bf \Gamma})}
{\cal H}({\bf \Gamma}) W_{{\cal H},q}({\bf \Gamma})
 \eta^\pm_p({\bf \Gamma})  .\nonumber
\end{eqnarray}
Obviously because the energy operator commutes
with the Maxwell-Boltzmann operator,
$\tilde  W_{{\cal H},q} = W_{{\cal H},q}$
and $ \tilde W_{{\cal H},p} = W_{{\cal H},p}$.

The Hamiltonian is the sum of
the kinetic  and potential energies,
which is indeed a linear combination of pure functions,
\begin{equation}
\hat {\cal H} = {\cal K}(\hat {\bf p}) + U(\hat{\bf q}) .
\end{equation}
Hence one can also write
\begin{eqnarray} \label{Eq:<H>,W}
\lefteqn{
\left< \hat {\cal H} \right>_{z,\beta,V}^\pm
}  \\
& = &
\left< \hat {\cal K} \right>_{z,\beta,V}^\pm
+
\left< \hat U \right>_{z,\beta,V}^\pm
\nonumber \\ & = &
\frac{1}{\Xi^\pm}
\sum_{N=0}^\infty \frac{z^N}{h^{dN} N!}
\int \mathrm{d}{\bf \Gamma}\;
e^{-\beta{\cal H}({\bf \Gamma})}
\left\{ {\cal K}({\bf p}) W_p({\bf \Gamma})  \eta^\pm_q({\bf \Gamma})
\right. \nonumber \\ && \left. \mbox{ }
+ U({\bf q}) W_q({\bf \Gamma})  \eta^\pm_p({\bf \Gamma})
\right\}
\nonumber \\ & = &
\frac{1}{\Xi^\pm}\sum_{N=0}^\infty \frac{z^N}{h^{dN} N!}
\int \mathrm{d}{\bf \Gamma}\;
e^{-\beta{\cal H}({\bf \Gamma})}
 {\cal H}({\bf \Gamma}) W_p({\bf \Gamma})  \eta^\pm_q({\bf \Gamma})
\nonumber \\ & = &
\frac{1}{\Xi^\pm}\sum_{N=0}^\infty \frac{z^N}{h^{dN} N!}
\int \mathrm{d}{\bf \Gamma}\;
e^{-\beta{\cal H}({\bf \Gamma})}
 {\cal H}({\bf \Gamma}) W_q({\bf \Gamma})  \eta^\pm_p({\bf \Gamma}) .\nonumber
\end{eqnarray}
The penultimate equality follows by taking the complex conjugate
of the average potential energy in the second equality.
The final equality follows by taking the complex conjugate
of the average kinetic  energy in the second equality.
One sees from these that one can obtain the average energy
using the same commutation function as for the partition function.

Differentiating with respect to $-\beta$ the defining equation
for $W_p$,
Eq.~(\ref{Eq:def-Wp}),
one obtains
\begin{eqnarray} \label{Eq:WHp}
{\cal H}({\bf \Gamma}) e^{-\beta{\cal H}({\bf \Gamma})}
\lefteqn{
W_{p}({\bf \Gamma})
-
e^{-\beta{\cal H}({\bf \Gamma})}
\frac{\partial  W_{p}({\bf \Gamma})}{\partial \beta }
} \nonumber \\
& = &
\frac{\langle{\bf q}| \hat{\cal H}  e^{-\beta\hat{\cal H}}|{\bf p}\rangle
}{\langle{\bf q}| {\bf p}\rangle }
\nonumber \\ & = &
e^{-\beta{\cal H}({\bf \Gamma})}
{\cal H}({\bf \Gamma})  W_{{\cal H},p}({\bf \Gamma}) ,
\end{eqnarray}
since $\tilde W_{{\cal H},p}=W_{{\cal H},p}$.
Hence $W_{{\cal H},p}= W_{p}
- {\cal H}^{-1} \partial  W_{p}/\partial \beta $.
Inserting the left hand side into Eq.~(\ref{Eq:<H>,WH}),
and comparing this with  Eq.~(\ref{Eq:<H>,W})
one sees that in order for the two expressions
for the average energy to be equal
one must have
\begin{equation}
\sum_{N=0}^\infty \frac{z^N}{h^{dN} N!}
\int \mathrm{d}{\bf \Gamma}\;
e^{-\beta{\cal H}({\bf \Gamma})}
\frac{\partial  W_{p}({\bf \Gamma})}{\partial \beta }
 \eta^\pm_q({\bf \Gamma})
= 0  .
\end{equation}
An analogous result holds for
$ {\partial  W_{q}({\bf \Gamma})}/{\partial \beta } $.

\subsubsection{A Useful Result}

I now prove that this vanishes individually
for each particle number $N$ and for each permutation $\hat{\mathrm P}$.
This turns out to be necessary for thermodynamic consistency,
as will be explained below.

With $|{\bf p}\rangle = V^{-N/2} e^{- {\bf p} \cdot {\bf r}/i\hbar}$,
one has that $\hat {\cal H} |{\bf p}\rangle
= {\cal H}({\bf p},{\bf r}) |{\bf p}\rangle$.
Also since the permutations are between  identical particles,
${\cal H}({\bf p},{\bf q})
={\cal H}(\hat{\mathrm P}{\bf p},{\bf q})
={\cal H}({\bf p},\hat{\mathrm P}{\bf q})$,
for any permutation $\hat{\mathrm P}$.
(This assumes that there are no velocity dependent forces,
which is the usual case. Lorentz forces need thought.)
Hence
\begin{equation}
 {\cal H}({\bf p}, {\bf q})
\langle \hat{\mathrm P} {\bf p} | {\bf q} \rangle
=
\langle \hat{\mathrm P} {\bf p} | \hat {\cal H} |{\bf q} \rangle .
\end{equation}
With this one has
\begin{eqnarray} \label{Eq:W'(eta)=0}
\lefteqn{
\sum_{N=0}^\infty \frac{z^N}{h^{dN} N!}
\int \mathrm{d}{\bf \Gamma}\;
e^{-\beta{\cal H}({\bf \Gamma})}
\frac{\partial  W_{p}({\bf \Gamma})}{\partial \beta }
 \eta^\pm_q({\bf \Gamma})
} \nonumber \\
& = &
\sum_{N=0}^\infty \frac{z^N}{h^{dN} N!}
\int \mathrm{d}{\bf \Gamma}\;
e^{-\beta{\cal H}({\bf \Gamma})}
 \eta^\pm_q({\bf \Gamma})
 \nonumber \\ && \mbox{ } \times
\left\{  {\cal H}({\bf \Gamma}) W_{{\cal H},p}({\bf \Gamma})
-  {\cal H}({\bf \Gamma}) W_{p}({\bf \Gamma}) \right\}
\nonumber \\ & = &
\sum_{N=0}^\infty \frac{z^N V^N}{h^{dN} N!}
\sum_{\hat{\mathrm P}} (\pm 1)^p
\int \mathrm{d}{\bf \Gamma}\;
\langle \hat{\mathrm P} {\bf p} | {\bf q} \rangle
\nonumber \\ && \mbox{ } \times
\left\{
\langle  {\bf q}
| e^{\beta \hat{\cal H}} \hat{\cal H} |
{\bf p} \rangle
-
{\cal H}({\bf p}, {\bf q})
\langle  {\bf q}
| e^{\beta \hat{\cal H}} |
{\bf p} \rangle
\right\}
\nonumber \\ & = &
\sum_{N=0}^\infty \frac{z^N V^N}{h^{dN} N!}
\sum_{\hat{\mathrm P}} (\pm 1)^p
\int \mathrm{d}{\bf \Gamma}
\nonumber \\ && \mbox{ } \times
\left\{
\langle \hat{\mathrm P} {\bf p} | {\bf q} \rangle \,
\langle  {\bf q}
| e^{\beta \hat{\cal H}} \hat{\cal H} |
{\bf p} \rangle
-
\langle \hat{\mathrm P} {\bf p}
| \hat{\cal H} | {\bf q} \rangle \,
\langle  {\bf q}
| e^{\beta \hat{\cal H}} |
{\bf p} \rangle
\right\}
\nonumber \\ & = &
\sum_{N=0}^\infty \frac{z^N}{h^{dN} N!}
\sum_{\hat{\mathrm P}} (\pm 1)^p
\int \mathrm{d}{\bf p}\;
\nonumber \\ && \mbox{ } \times
\left\{
\langle \hat{\mathrm P} {\bf p}
| e^{-\beta \hat{\cal H}} \hat{\cal H} |
 {\bf p} \rangle
 -
\langle \hat{\mathrm P} {\bf p}
|  \hat{\cal H} e^{-\beta \hat{\cal H}} |
 {\bf p} \rangle
 \right\}
\nonumber \\ & = &
0.
\end{eqnarray}
The term in braces
vanishes for each $N$ and for each permutation $\hat{\mathrm P}$,
as was to be shown.
This result will prove important shortly.
It also vanishes for each  momentum configuration ${\bf p}$,
which will not be required.

%
\section{Loop Expansion, Grand Potential, and Average Energy} \label{Sec:loop}
\setcounter{equation}{0} \setcounter{subsubsection}{0}
%

\subsection{Symmetrization Loops}

The symmetrization function can be written in terms of loops.
Any particular particle permutation operator
can be factored into loop permutation operators.
A loop is a connected series of pair transpositions.
Hence the sum over all permutation operators
can be written as the sum over all possible factors of loop permutations,
\begin{eqnarray}
\sum_{\hat{\mathrm P} } (\pm1)^p\; \hat{\mathrm P}
& = &
\hat {\mathrm I}
\pm \sum_{i,j} \!' \; \hat{\mathrm P}_{ij}
+ \sum_{i,j,k} \!' \; \hat{\mathrm P}_{ij} \hat{\mathrm P}_{jk}
\nonumber \\ & & \mbox{ }
+ \sum_{i,j,k,l} \!\!' \; \hat{\mathrm P}_{ij} \hat{\mathrm P}_{kl}
\pm \ldots
\end{eqnarray}
Here $\hat{\mathrm P}_{jk}$ is the transpose of particles $j$ and $k$.
The prime on the sums restrict them to unique loops,
with each index being different.
The first term is just the identity.
The second term is a dimer loop,
the third term is a trimer loop,
and the fourth term shown is the product of two different dimers.

With this,
the symmetrization function,
$\eta^\pm_q({\bf \Gamma})$ $= \sum_{\hat{\mathrm P} }(\pm 1)^p$
$ \langle \hat{\mathrm P}{\bf p} | {\bf q} \rangle
/\langle {\bf p} | {\bf q} \rangle$,
is the sum of the expectation values of these loops.
The monomer symmetrization function is obviously unity,
$ \eta^{(1)}_q
\equiv
{ \langle {\bf p} | {\bf q} \rangle }/{\langle {{\bf p}} | {\bf q} \rangle}
= 1$.

The dimer  symmetrization factor in the microstate ${\bf \Gamma}$
for particles $j$ and $k$ is
\begin{eqnarray} \label{Eq:eta^(2)}
\eta^{\pm(2)}_{q;jk}
& = &
\frac{ \pm \langle {\hat{\mathrm P}_{jk}{\bf p}}
| {\bf q} \rangle
}{\langle {{\bf p}} | {\bf q} \rangle}
\nonumber \\ & = &
\frac{ \pm
\langle {{\bf p}_k}| {{\bf q}_j} \rangle
\langle {{\bf p}_j} | {{\bf q}_k} \rangle
}{
\langle {{\bf p}_j}| {{\bf q}_j} \rangle
\langle {{\bf p}_k} | {{\bf q}_k} \rangle }
\nonumber \\ & = &
\pm
e^{ ({\bf q}_k-{\bf q}_j) \cdot {\bf p}_j /i\hbar }
e^{ ({\bf q}_j-{\bf q}_k) \cdot {\bf p}_k /i\hbar } .
\end{eqnarray}
Recall that
$|{\bf p}\rangle \equiv
{e^{-{\bf p}\cdot{\bf r}/i \hbar}}/{V^{N/2}}$.
Note that since the basis functions are the product
of single particle functions,
the expectation value factorizes
leaving only the permuted particles to contribute.

Similarly the trimer symmetrization factor for particles $j$, $k$, and $ l$ is
\begin{eqnarray}
\eta^{\pm(3)}_{q;jkl}
& = &
\frac{ \langle {\hat{\mathrm P}_{jk}\hat{\mathrm P}_{kl}{\bf p}}
| {\bf q}  \rangle
}{
\langle {{\bf p}} | {\bf q}  \rangle }
 \\ & = &
\frac{
\langle {{\bf p}_k}| {{\bf q}_j} \rangle
\langle {{\bf p}_j} | {{\bf q}_l} \rangle
\langle {{\bf p}_l} | {{\bf q}_k} \rangle
}{
\langle {{\bf p}_j}| {{\bf q}_j} \rangle
\langle {{\bf p}_k} | {{\bf q}_k} \rangle
\langle {{\bf p}_l} | {{\bf q}_l} \rangle}
\nonumber \\ & = &
e^{ ({\bf q}_j-{\bf q}_k) \cdot {\bf p}_k /i\hbar }
e^{ ({\bf q}_k-{\bf q}_l) \cdot {\bf p}_l /i\hbar }
e^{ ({\bf q}_l-{\bf q}_j) \cdot {\bf p}_j /i\hbar } . \nonumber
\end{eqnarray}

In general
the $l$-loop symmetrization factor is
\begin{equation} \label{Eq:tilde-eta-l}
\eta_{q;1 \ldots l}^{\pm(l)}
=
(\pm 1)^{l-1}
e^{ {\bf q}_{l1} \cdot {\bf p}_1 /i\hbar }
\prod_{j=1}^{l-1}
e^{ {\bf q}_{j+1,j} \cdot {\bf p}_j /i\hbar } ,
\end{equation}
where $ {\bf q}_{jk} \equiv {\bf q}_{j}- {\bf q}_{k}$.
Recall that $\eta_p = \eta_q^*$.

The product of Fourier exponentials
that occurs here makes each specific $l$-loop symmetrization factor
highly oscillatory unless successive particles
around the loop are close together
in both momentum and position space.
This means that the only non-zero contributions to phase space integrals
come from such compact loops,
since otherwise their oscillations would average to zero.
(See Eq.~(\ref{Eq:dimer-config}) below for an explicit example.)

With these  symmetrization factors,
the symmetrization function can be written as a series of loop products,
\begin{eqnarray}
\eta^\pm_q({\bf \Gamma})
& = &
1
+ \sum_{ij}\!'  \eta_{q;ij}^{\pm(2)}
+ \sum_{ijk}\!'  \eta_{q;ijk}^{\pm(3)}
\nonumber \\ &&  \mbox{ }
+ \sum_{ijkl}\!' \eta_{q;ij}^{\pm(2)}
\eta_{q;kl}^{\pm(2)}
+ \ldots
\end{eqnarray}
Here the superscript is the order of the loop,
and the subscripts are the atoms involved in the loop.
The prime signifies that the sum is over unique loops
(ie.\ each configuration of particles in loops occurs once only)
with each index different
(ie.\ no particle may belong to more than one loop).

In this
one can identify the terms with only a single loop,
and define the single loop symmetrization function,
\begin{eqnarray} \label{Eq:dot-eta}
\dot\eta_q^\pm({\bf \Gamma})
& \equiv &
\sum_{jk}\!'  \eta_{q;jk}^{\pm(2)}
+ \sum_{jkm}\!'  \eta_{q;jkm}^{\pm(3)}
+ \sum_{jkmn}\!\!' \; \eta_{q;jkmn}^{\pm(4)}
+ \ldots
\nonumber \\ & \equiv &
\sum_{l=2}^\infty \dot\eta_{q}^{\pm(l)}({\bf \Gamma}) .
\end{eqnarray}
This gives the  single $l$-loop symmetrization function
$ \dot\eta_{q}^{\pm(l)}({\bf \Gamma})$
as the sum over the $N!/(N-l)!l$ distinct arrangements
of the $l$ particle labels.
In taking the average of each single loop symmetrization function,
since the particles are identical,
all the terms in the sum over particles gives the same average value.
Therefore, what will be shown next to be the loop grand potential
can be obtained by evaluating any one arrangement,
say particles $1,2,\ldots,l$,
\begin{eqnarray}
-\beta \Omega_{W_p}^{\pm(l)}
& \equiv &
\left\langle \dot\eta_{q}^{\pm(l)}({\bf \Gamma})  \right\rangle_{W_p,1}
\nonumber \\ & = &
\left\langle
\sum_{j_1 \ldots j_l}\!\!' \; \eta_{q;j_1 \ldots j_l}^{\pm(l)}
\right\rangle_{W_p,1}
\nonumber \\ & = &
\left\langle
\frac{N!}{(N-l)!l} \, \eta_{q;1\ldots l}^{\pm(l)}
\right\rangle_{W_p,1} .
\end{eqnarray}

\subsection{Grand Potential} \label{Sec:Omega}

An expression for the grand potential is now derived
that invokes a factorized form for integrals
involving the symmetrization function.
This factorization ansatz is exact for  non-interacting systems,
as has been explicitly confirmed for the quantum ideal gas\cite{STD2}
and for a system of non-interacting quantum harmonic oscillators.
\cite{Attard18}
The ansatz neglects correlations between loops in the interacting case,
and in this case it is expected to be exact in the thermodynamic limit,
as is discussed in \S \ref{Sec:Ex-dimer-corr} below.
That large volume limit is taken at constant fugacity or density;
it may well be that  for interacting particles
the factorization ansatz breaks down
in the constant volume, high density limit.

In fact the factorization ansatz is not essential for the theory,
since one can just take the grand potential from the logarithm of
the partition function,
the latter having been evaluated using the full series
for the symmetrization function,
and similarly for any average.
Undoubtedly, however,
it would be somewhat tedious to calculate the series in full,
and the exponential that results from the factorization
is convenient and much more rapidly converging.

The monomer grand partition function
comes from setting $\eta^\pm=1$,
\begin{equation}
\Xi_{W_p,1}^\pm =
\sum_{N=0}^\infty \frac{z^N}{h^{dN}N!}
\int \mathrm{d}{\bf p}\,\mathrm{d}{\bf q}\;
e^{-\beta {\cal H}({\bf p},{\bf q})}
W_p({\bf p},{\bf q}) .
\end{equation}
Recall that the imaginary part of $W_p$ is odd in momentum,
and that $W_p^* = W_q$,
and so the subscript $p$ or $q$ (and also the superscript $\pm$)
is redundant in the monomer case.
Nevertheless it seems best to signify explicitly
which of these commutation function is being used.

The ratio of the full to the monomer grand partition function
is the monomer average of the symmetrization function,
\begin{eqnarray}
\frac{\Xi^\pm_{W_p}}{\Xi_{W_p,1}}
& = &
\left<\eta_q^\pm \right>_{W_p,1}
\nonumber \\ & = &
1
+ \left< \sum_{ij}\!'  \eta_{q;ij}^{\pm(2)} \right>_{W_p,1}
+ \left< \sum_{ijk}\!'   \eta_{q;ijk}^{\pm(3)}\right>_{W_p,1}
\nonumber \\ &&  \mbox{ }
+ \left< \sum_{ijkl}\!'  \eta_{q;ij}^{\pm(2)}
\eta_{q;kl}^{\pm(2)} \right>_{W_p,1}
+ \ldots
\nonumber \\ & = &
1
+ \left< \frac{N}{(N-2)!2}  \eta_{q;12}^{\pm(2)} \right>_{W_p,1}
\nonumber \\ &&  \mbox{ }
+   \left< \frac{N!}{(N-3)!3}  \eta_{q;123}^{\pm(3)}\right>_{W_p,1}
\nonumber \\ &&  \mbox{ }
+ \frac{1}{2}
\left<  \frac{N!}{(N-2)!2} \eta_{q;12}^{\pm(2)} \right>_{W_p,1}^2
+ \ldots
\nonumber \\ & = &
\sum_{\{m_l\}}
\frac{1}{m_l!}
\prod_{l=2}^\infty
\left< \frac{N!}{(N-l)!l} \eta_{q;1\ldots l}^{\pm(l)} \right>^{m_l}_{W_p,1}
\nonumber \\ & = &
\prod_{l=2}^\infty
\sum_{m_l=0}^\infty
\frac{1}{m_l!}
\left< \frac{N!}{(N-l)!l}  \eta_{q;1\ldots l}^{\pm(l)} \right>^{m_l}_{W_p,1}
\nonumber \\ & = &
\prod_{l=2}^\infty
e^{ -\beta \Omega_{W_p}^{\pm(l)}}.
\end{eqnarray}
The third and following equalities write the average of the product
as the product of the averages.
This is valid in the thermodynamic limit,
since the product of the average of two loops scales as $V^2$,
whereas the correlated interaction of two loops scales as $V$,
and similarly for all the other products.
As mentioned above, only compact loops contribute to the integral;
see also \S \ref{Sec:Ex-dimer-corr} below.
The combinatorial factor accounts for the number of unique loops
in each term.

The grand potential is the logarithm of the partition function
$\Omega^\pm = -k_\mathrm{B}T \ln \Xi^\pm$.
The monomer grand potential is given by
$-\beta\Omega_{W_p,1}^\pm = \ln \Xi_{W_p,1}^\pm$.
For the monomer term
the superscript $\pm$  is redundant and will usually be dropped.

The difference between the full grand potential
and the monomer grand potential is just
the series of loop grand potentials,
\begin{eqnarray} \label{Eq:Omega-l}
- \beta [\Omega^\pm_{W_p} - \Omega_{W_p,1} ]
& = &
\ln \frac{\Xi^\pm_{W_p}}{\Xi_{W_p,1}}
\nonumber \\ & = &
\sum_{l=2}^\infty
\left< \frac{N!}{(N-l)!l} \eta_{q;1\ldots l}^{\pm(l)} \right>_{W_p,1}
\nonumber \\ & \equiv &
-\beta  \sum_{l=2}^\infty \Omega_{W_p}^{\pm(l)} .
\end{eqnarray}
Here  $\eta_q$ is paired with $W_p$;
one could alternatively pair $\eta_p$ and $W_q$.
Since these give the same, real, value for  the grand potential,
it is redundant and somewhat pedantic to attach the $W_p$ subscript.

\subsubsection{Excess Dimer Correlation} \label{Sec:Ex-dimer-corr}

Above,
the product of two dimers was written as
\begin{equation}
\left< \sum_{jkmn}\!\!' \;  \eta_{q;jk}^{\pm(2)}
\eta_{q;mn}^{\pm(2)} \right>_{W_p,1}
=
\frac{1}{2}
\left<  \frac{N!}{(N-2)!2} \eta_{q;jk}^{\pm(2)} \right>_{W_p,1}^2 .
\end{equation}
The prime indicates that each configuration of loops occurs once only,
and that no two indeces are equal.

\comment{ 
Evaluation of the ratio of the two sides of this
as a function of temperature and density
would give a guide to the quantitative accuracy
of the factorization ansatz for a system of  interacting particles.
Initially, one could do this in the classical limit, $W_p=1$.

For a configuration ${\bf \Gamma}$,
one can identify and store the pairs such that
$r_{jk} < f_\mathrm{cut} \Lambda$.
One can give them a one-dimensional label, $J \Leftrightarrow \{jk\}$.
One need only use these pairs to evaluate the left hand side above,
which is a great saving.
If no index appears more than once in the list,
one can simply square the sum,
$\sum_{J<J'}  \eta_{q;J}^{\pm(2)} \eta_{q;J'}^{\pm(2)}
= [\sum_{J}  \eta_{q;J}^{\pm(2)}]^2/2
- \sum_{J}  (\eta_{q;J}^{\pm(2)})^2/2$.
} 

With $\eta_{q;jk}^{\pm(2)} = e^{-{\bf p}_{jk} \cdot {\bf q}_{jk} /i\hbar}$,
in the case of a classical average (ie.\ $\langle \ldots \rangle_{1,1}$),
one can perform the momentum integrals by completing the square,
\begin{eqnarray} \label{Eq:dimer-config}
\lefteqn{
\left\langle \eta_{q;jk}^{\pm(2)} \right\rangle_{1,1}
} \nonumber \\
& = &
\frac{\pm Z^{-1}}{h^{3N}N!}
\int \mathrm{d}{\bf \Gamma} \;
e^{-\beta {\cal H}({\bf \Gamma})} e^{-{\bf p}_{jk} \cdot {\bf q}_{jk} /i\hbar}
\nonumber \\ & = &
\frac{\pm Q^{-1}}{V^{N}N!}
\int \mathrm{d}{\bf q}\;
e^{-\beta U({\bf q})}
e^{-m q^2_{jk}/2\beta\hbar^2}
e^{-m q^2_{kj}/2\beta\hbar^2}
\nonumber \\ & = &
\frac{\pm Q^{-1}}{V^{N}N!}
\int \mathrm{d}{\bf q}\;
e^{-\beta U({\bf q})}
e^{-\pi q^2_{jk}/\Lambda^2}
e^{-\pi q^2_{kj}/\Lambda^2} ,
\end{eqnarray}
where the thermal wavelength is
$\Lambda \equiv \sqrt{2\pi\hbar^2\beta/m}$.
The integrand is zero except for separations less than the thermal wavelength,
$q_{jk} \alt \Lambda$.
This means that effectively particle $j$ is tied to particle $k$,
which means that a volume integral is lost,
so that
$\left\langle \eta_{q;jk}^{\pm(2)} \right\rangle_{1,1}
\sim  {\cal O}(V^{-1}) $.

It is clear from this that
$\langle\sum_{jk}\!\!' \;  \eta_{q;jk}^{\pm(2)} \rangle_{W_p,1}$
involves an integral over position space
of the two particle density $\rho^{(2)}$.
Recall that in classical statistical mechanics,
$\rho^{(2)}$at large separations
goes asymptotically  as the square of the singlet densities.
Similarly,
$\langle\sum_{jkmn}\!\!' \;  \eta_{q;jk}^{\pm(2)}
\eta_{q;mn}^{\pm(2)} \rangle_{W_p,1}$
involves an integral of the four particle density $\rho^{(4)}$,
which for large separations between the pairs goes like
$\rho^{(4)}({\bf q}_{j},{\bf q}_{k},{\bf q}_{m},{\bf q}_{n}) \sim
\rho^{(2)}({q}_{jk}) \rho^{(2)}({q}_{mn})$.
Hence in the thermodynamic limit,
the relative error in the factorization of the dimer product
vanishes
\begin{eqnarray}
\frac{
\left< \sum_{jkmn}\!\!' \;  \eta_{q;jk}^{\pm(2)}
\eta_{q;mn}^{\pm(2)} \right>_{W_p,1}
-
\frac{1}{2} \left<  \sum_{jk}\!\!' \;  \eta_{q;jk}^{\pm(2)} \right>_{W_p,1}^2
}{
\frac{1}{2}
\left< \sum_{jk}\!\!' \;  \eta_{q;jk}^{\pm(2)}\right>_{W_p,1}^2
}
& \rightarrow & 0
\nonumber \\  \mbox{ }
z,\,T =\mbox{const.}, \;
V \rightarrow \infty.
\end{eqnarray}
The denominator goes like $ {\cal O}(V^{-2}) $.
The numerator goes like $ {\cal O}(V^{-3})$,
because three volume integrals are lost
when all four particles are correlated,
which is required for the difference to be non-zero.
An analogous argument holds for the arbitrary product of arbitrary loops.
This justifies the factorization of the symmetrization
function in the thermodynamic limit.

\subsection{Energy} \label{Sec:Av-H-Omega}

The most likely energy can be written as the temperature derivative
of the grand potential.
As above the latter can be written as a series of loop derivatives.
One has
\begin{equation}
\overline E_{W_p}
=
\frac{\partial \beta \Omega_{W_p}}{\partial \beta }
=
\sum_{l=1}^\infty \overline E_{W_p,l} .
\end{equation}
The monomer term is
\begin{eqnarray}  \label{Eq:olE1}
\overline E_{W_p,1} & = &
\frac{\partial \beta \Omega_{W_p,1}}{\partial \beta }
=
\frac{-\partial \ln \Xi_{W_p,1}}{\partial \beta }
\nonumber \\ & = &
\frac{1}{ \Xi_{W_p,1} }
\sum_{N=0}^\infty \frac{z^N}{h^{dN}N!}
\int \mathrm{d}{\bf \Gamma}\;
e^{-\beta {\cal H}({\bf \Gamma})}
\nonumber \\ &  & \mbox{ } \times
\left\{ {\cal H}({\bf \Gamma}) W_{p}({\bf \Gamma})
- \frac{ \partial  W_{p}({\bf \Gamma})}{\partial \beta} \right\}
\nonumber \\ & = &
\left< {\cal H} \,  \right>_{W_p,1}
\nonumber \\ & = &
\left< {\cal H} \,  \right>_{W_{{\cal H},p},1} .
\end{eqnarray}
In and about Eq.~(\ref{Eq:WHp}) above,
it was shown that
$W_{{\cal H},p} = W_{p}
- {\cal H}^{-1} \partial  W_{p}/\partial \beta $.
It was also shown in Eq.~(\ref{Eq:W'(eta)=0})
that the integral of the second part must vanish
\emph{for each $N$ and for each permutation $\hat{\mathrm P}$}.
One can therefore interchange $W_{{\cal H},p} $ and $ W_{p} $
in the present monomer integrals, including
 $\Xi_{W_p,1} =\Xi_{W_{{\cal H},p},1} $.
This says that $\overline E_{W_p,1} = \overline E_{W_{{\cal H},p},1}$,
so that the commutation function in the subscript is redundant.

For the loops $l \ge 2$ one has
\begin{eqnarray} \label{Eq:olEl}
\overline E_{W_p,l}
& = &
\frac{\partial \beta \Omega_{W_p}^{\pm(l)}}{\partial \beta }
\nonumber \\ & = &
\frac{-\partial }{\partial \beta }
\left< \frac{N!}{(N-l)!l} \eta_{q,1\ldots l}^{\pm(l)} \right>_{W_p,1}
\nonumber \\ & = &
-\beta \Omega_{W_p}^{\pm(l)}
\frac{1}{\Xi_{W_p,1}} \frac{\partial \Xi_{W_p,1}}{\partial \beta }
\nonumber \\ &  & \mbox{ }
+
\frac{1}{ \Xi_{W_p,1} }
\sum_{N=l}^\infty \frac{z^Nh^{-dN}}{(N-l)!l}
\int \mathrm{d}{\bf \Gamma}\;
e^{-\beta {\cal H}({\bf \Gamma})}
\nonumber \\ &  & \mbox{ } \times
\left\{ {\cal H}({\bf \Gamma}) W_{p}({\bf \Gamma})
- \frac{ \partial  W_{p}({\bf \Gamma})}{\partial \beta} \right\}
 \eta_{q,1\ldots l}^{\pm(l)}
\nonumber \\ & = &
\beta \Omega_{W_p}^{\pm(l)}\, \overline E_{W_p,1}
+
\left< {\cal H} \, \dot\eta_{q}^{\pm(l)} \right>_{W_p,1}
\nonumber \\ & = &
\beta \Omega_{W_p}^{\pm(l)} \, \overline E_{W_{{\cal H},p},1}
+
\left< {\cal H} \, \dot\eta_{q}^{\pm(l)}
\right>_{W_{{\cal H},p},1}  .
\end{eqnarray}
Recall that
$ \dot\eta_{q}^{\pm(l)}({\bf \Gamma})
= \sum_{j_1 \ldots j_l}\!\!\!\!\!\!\!\!\!\!' \;\;\;\;
 \eta_{q;j_1 \ldots j_l}^{\pm(l)} $.
Again, Eq.~(\ref{Eq:W'(eta)=0})
has been applied here,
showing that one can equally use
$W_{{\cal H},p} $ or $ W_{p} $.
This expression has the form of the average of a fluctuation,
\begin{equation} \label{Eq:olEl-fluctn}
\overline E_{W_p,l}
=
\left<  \left[ {\cal H}  - \overline E_{W_p,1} \right]
\left[ \dot\eta_{q}^{\pm(l)}
- \langle  \dot\eta_{q}^{\pm(l)} \rangle_{W_p,1}
\right]  \right>_{W_p,1} ,
\end{equation}
where $l \ge 2$,
and where either commutation function may be used.
This result will be derived by a different route in
\S \ref{Sec:dens-H} below.

\subsubsection{Heat Capacity}

The heat capacity at constant volume and fugacity is
\begin{eqnarray}
C_V
& = &
\frac{\partial \overline E}{ \partial T}
\nonumber \\ & = &
-k_\mathrm{B} \beta^{2}
\frac{\partial \overline E}{ \partial \beta}
\nonumber \\ & = &
-k_\mathrm{B} \beta^{2}
\sum_{l=1}^\infty
\frac{\partial \overline E_l }{ \partial \beta}
.
\end{eqnarray}

The monomer contribution comes from ($W_p$ is not constant)
\begin{eqnarray}
\frac{\partial \overline E_1 }{ \partial \beta}
& = &
\frac{\partial  }{ \partial \beta}
\frac{1}{ \Xi(W_{p}, 1) }
\sum_{N=0}^\infty \frac{z^N}{h^{dN}N!}
\int \mathrm{d}{\bf \Gamma}\;
\nonumber \\ & & \mbox{ } \times
e^{-\beta {\cal H}({\bf \Gamma})} {\cal H}({\bf \Gamma})
W_{p}({\bf \Gamma})
\nonumber \\ & = &
\left<  {\cal H} -
\frac{\partial \ln W_p }{ \partial \beta}
\right>_{W_p,1} \overline E_1
\nonumber \\ & & \mbox{ }
- \left<
{\cal H}^2 - {\cal H} \frac{\partial \ln W_p }{ \partial \beta}
\right>_{W_p,1}
\nonumber \\ & = &
\overline E_1^2
- \left<
{\cal H}^2 - {\cal H} \frac{\partial \ln W_p }{ \partial \beta}
\right>_{W_p,1} .
\end{eqnarray}

The loop contribution $l\ge 2$
comes from the derivative of Eq.~(\ref{Eq:olEl}),
with $W_{{\cal H},p} \Rightarrow W_{p}$,
\begin{eqnarray}
\frac{\partial \overline E_l }{ \partial \beta}
& = &
\frac{\partial  }{ \partial \beta}
\left\{
\beta \Omega_l\,\overline E_1
+
\frac{1}{ \Xi_{1p}^\pm }
\sum_{N=l}^\infty \frac{z^Nh^{-dN}}{(N-l)!l}
\int \mathrm{d}{\bf \Gamma}\;
\right. \nonumber \\ &  & \left. \mbox{ } \times
e^{-\beta {\cal H}({\bf \Gamma})}
{\cal H}({\bf \Gamma}) W_{p}({\bf \Gamma})
 \eta_{q;1 \ldots l}^{\pm(l)}
\rule{0cm}{0.6cm} \right\}
\nonumber \\ & = &
\overline E_l \overline E_1
+ \beta \Omega_l
\frac{\partial \overline E_1 }{ \partial \beta}
+ \overline E_1 \left[ \overline E_l - \beta \Omega_l \overline E_1 \right]
\nonumber \\ &  & \mbox{ }
-
\left< {\cal H}^2
- {\cal H} \frac{\partial \ln W_{p}}{\partial \beta}
\right>_{W_p, \dot\eta_{q}^{\pm(l)}} .
\end{eqnarray}

\comment{ 
For what its worth,
the second derivative with respect to $\beta$
of the defining equation
for the grand partition function commutation function,
Eq.~(\ref{Eq:def-Wp}),
gives
\begin{eqnarray}
\lefteqn{
e^{-\beta{\cal H}({\bf \Gamma})}
{\cal H}({\bf \Gamma})^2  W_{{\cal H}^2,p}({\bf \Gamma})
} \nonumber \\
& = &
\frac{\langle{\bf q}| \hat{\cal H}^2  e^{-\beta\hat{\cal H}}|{\bf p}\rangle
}{\langle{\bf q}| {\bf p}\rangle }
\nonumber \\ & = &
{\cal H}({\bf \Gamma})^2 e^{-\beta{\cal H}({\bf \Gamma})}
W_{p}({\bf \Gamma})
- 2 {\cal H}({\bf \Gamma})
e^{-\beta{\cal H}({\bf \Gamma})}
\frac{\partial  W_{p}({\bf \Gamma})}{\partial \beta }
\nonumber \\ &  & \mbox{ }
+
e^{-\beta{\cal H}({\bf \Gamma})}
\frac{\partial^2  W_{p}({\bf \Gamma})}{\partial \beta^2 } .
\end{eqnarray}
The utility of this is unclear.
} 

%
\section{Multi-Particle Density} \label{Sec:rho}
\setcounter{equation}{0} \setcounter{subsubsection}{0}
%

Consider a position configuration of $n$ particles,
${\bf Q}^n = \{{\bf Q}_1,{\bf Q}_2,\ldots,{\bf Q}_n\}$.
The distinct $n$-particle density quantum operator for this is
\begin{equation}
\hat \rho^{(n)}({\bf Q}^n;{\bf r})
=
\sum_{k_1,\ldots,k_n}^N \!\!\!\!\!\!''\;\;
\prod_{j=1}^n \delta({\bf Q}_j - {\bf r}_{k_j}).
\end{equation}
The sum is over the particle indeces,
and the double prime indicates that in any term no two indeces are equal.
If there are $N$ particles in the system,
then there are $N!/(N-n)!$ terms in the sum.
This says any particle is at  ${\bf Q}_1$,
any different particle is at  ${\bf Q}_2$ etc.
The density operator is evidently unchanged by a permutation
of the ${\bf Q}_j $ or of the particle labels.

Since this is only a function of the position operator,
the commutation function $W_q$ or $W_p$ may be used for its average.
The average $n$-particle density is
\begin{eqnarray}
\lefteqn{
\rho^{(n)}_{W_p}({\bf Q}^n)
} \nonumber \\
& \equiv &
\left\langle \hat \rho^{(n)}({\bf Q}^n;{\bf q}) \right\rangle_{W_p,\eta^\pm_q}
\nonumber \\ & = &
\frac{1}{ \Xi_{W,\eta^\pm} }
\sum_{N=0}^\infty \frac{z^N}{h^{dN} N!}
\int \mathrm{d}{\bf \Gamma}\;
e^{-\beta{\cal H}({\bf \Gamma})}
W_p({\bf \Gamma}) \eta_q^\pm({\bf \Gamma})
\nonumber \\ && \mbox{ }\times
\sum_{k_1,\ldots,k_n}^N \!\!\!\!\!\!''\;\;
\prod_{j=1}^n \delta({\bf Q}_j - {\bf q}_{k_j}) .
\end{eqnarray}
It is worth including the subscript $W_p$ on the density
because there may be occasion when one requires a density
defined with respect to a commutation function
that is specific for a particular quantity that is to be averaged.
The subscript on the average, $W_p,\eta^\pm_q$,
could be replaced by, or augmented with, the subscript $z,V,T$.

It is worth mentioning that one does not actually
have to define the quantum operator density,
$\hat \rho^{(n)}({\bf Q}^n;{\bf q})$,
but that instead one can introduce the phase function
$\rho^{(n)}_{W_p}({\bf Q}^n)$ directly via the phase space integral.
More generally one can define the $n$-particle density
for positions and momenta in phase space as
\begin{equation}
\rho^{(n)}({\bm \gamma}^n;{\bf \Gamma})
=
\sum_{k_1,\ldots,k_n}^N \!\!\!\!\!\!''\;\;
\prod_{j=1}^n \delta({\bm \gamma}_j - {\bf \Gamma}_{k_j}),
\end{equation}
and its average as
\begin{eqnarray} \label{Eq:rho-n-gamma}
\lefteqn{
\rho^{(n)}_{W_p}({\bm \gamma}^n)
} \nonumber \\
& \equiv &
\frac{1}{ \Xi_{W,\eta^\pm} }
\sum_{N=0}^\infty \frac{z^N}{h^{dN} N!}
\int \mathrm{d}{\bf \Gamma}\;
e^{-\beta{\cal H}({\bf \Gamma})}
W_p({\bf \Gamma}) \eta_q^\pm({\bf \Gamma})
\nonumber \\ && \mbox{ }\times
\sum_{k_1,\ldots,k_n}^N \!\!\!\!\!\!''\;\;
\prod_{j=1}^n \delta({\bm \gamma}_j - {\bf \Gamma}_{k_j}) .
\end{eqnarray}
I shall use this below for the average energy.

The $n$-particle density for positions is evidently normalized as
\begin{equation}
\int \mathrm{d}{\bf Q}^n \; \rho_{W_p}^{(n)}({\bf Q}^n)
=
\left\langle \frac{N!}{(N-n)!} \right\rangle_{z,V,T} ,
\end{equation}
which is in agreement with the classical convention.\cite{Pathria72,TDSM}
One also has the reduction condition
\begin{eqnarray}
\lefteqn{
\int \mathrm{d}{\bf Q}_n \; \rho_{W_p}^{(n)}({\bf Q}^n)
} \nonumber \\
& = &
\!
\left\langle \frac{N!}{(N-n)!}  \right\rangle_{z,V,T}\!
\left\langle  \frac{N!}{(N-n+1)!} \right\rangle_{z,V,T}^{-1}
\rho_{W_p}^{(n-1)}({\bf Q}^{n-1})
\nonumber \\ & \approx &
(\overline N-n+1 )\; \rho_{W_p}^{(n-1)}({\bf Q}^{n-1}).
\end{eqnarray}

This  $n$-particle density can be used for the average of any operator
that is a function of the position operator.
For example,
suppose that the potential energy is the sum of pair potentials,
\begin{equation}
U({\bf q})
=
\frac{1}{2} \sum_{j,k}\!''\, u({\bf q}_j,{\bf q}_{k})
= \sum_{j,k}\!'\, u({\bf q}_j,{\bf q}_{k}).
\end{equation}
The double prime again indicates that $j \ne k$,
and the single prime indicates that the sum is over unique pairs, $j<k$.
The average potential energy in this case is
\begin{equation}
\left\langle \hat U  \right\rangle_{z,V,T}
=
\frac{1}{2}
\int \mathrm{d}{\bf Q}_1 \, \mathrm{d}{\bf Q}_2 \;
\rho^{(2)}_{W_p}({\bf Q}_1,{\bf Q}_2)\, u({\bf Q}_1,{\bf Q}_{2}) ,
\end{equation}
as can be confirmed by substituting in the definition of the pair density
and interchanging the order of integrations.
The factor of one half again corrects
for double counting of each particle pair.

Define the density fluctuation
\begin{eqnarray}
\Delta \hat\rho^{(n)}({\bf Q}^n;{\bf q})
& \equiv &
\hat\rho^{(n)}({\bf Q}^n;{\bf q})
-
\langle \hat\rho^{(n)}({\bf Q}^n;{\bf q}) \rangle_{W_p,1}
\nonumber \\ & \equiv &
\hat\rho^{(n)}({\bf Q}^n;{\bf q})
- \rho^{(n)}_{W_p,1}({\bf Q}^n) ,
\end{eqnarray}
and similarly for the symmetrization function.
With these the average density may be written
\begin{eqnarray} \label{Eq:rho-l}
\lefteqn{
\rho^{(n)}_{W_p}({\bf Q}^n)
} \nonumber \\
& \equiv &
\left\langle \hat \rho^{(n)}({\bf Q}^n;{\bf q}) \right\rangle_{W_p,\eta_q^\pm}
\nonumber \\ & = &
\frac{\Xi_{W_p,1}}{\Xi_{W_p,\eta_q^\pm}}
\left\langle \hat \rho^{(n)}({\bf Q}^n;{\bf q})
\, \eta_q^\pm({\bf \Gamma}) \right\rangle_{W_p,1}
\nonumber \\ & = &
\left\langle \eta_q^\pm({\bf \Gamma}) \right\rangle_{W_p,1}^{-1}
\left\{
\left\langle \hat \rho^{(n)}({\bf Q}^n;{\bf q}) \right\rangle_{W_p,1}
\left\langle \eta_q^\pm({\bf \Gamma}) \right\rangle_{W_p,1}
\right. \nonumber \\ && \left. \mbox{ }
+
\left\langle \Delta \hat \rho^{(n)}({\bf Q}^n;{\bf q})
\Delta \eta_q^\pm({\bf \Gamma})  \right\rangle_{W_p,1}
\right\}
\nonumber \\ & = &
\rho^{(n)}_{W_p,1}({\bf Q}^n)
+
\left\langle \Delta \hat \rho^{(n)}({\bf Q}^n;{\bf q})\,
\Delta \dot\eta_q^\pm({\bf \Gamma})   \right\rangle_{W_p,1}  .
\end{eqnarray}
This holds in the thermodynamic limit, $V\rightarrow \infty$,
$z,T =$ const.
Recall that $ \dot \eta_q^\pm $,
is the series of single loop symmetrization functions
defined in Eq.~(\ref{Eq:dot-eta}).
From Eq.~(\ref{Eq:Omega-l}), the prefactor in the second equality is
${\Xi_{W_p,1}}/{\Xi_{W_p,\eta_q^\pm}}
=
\langle \eta_q^\pm \rangle_{W_p,1}^{-1}
= e^{- \langle \dot \eta_q^\pm \rangle_{W_p,1} } $.
Also, since $\langle \Delta \hat \rho^{(n)}  \rangle_{W_p,1} = 0$,
one can write successively
\begin{eqnarray} \label{Eq:exp-dot-eta}
\langle \Delta \hat \rho^{(n)} \Delta\eta_q^\pm  \rangle_{W_p,1}
& = &
\langle \Delta \hat \rho^{(n)} \eta_q^\pm  \rangle_{W_p,1}
\nonumber \\ & = &
\langle \Delta \hat \rho^{(n)} [\eta_q^\pm-1] \rangle_{W_p,1}
\nonumber \\ & = &
\langle \Delta \hat \rho^{(n)} \dot\eta_q^\pm  \rangle_{W_p,1}
e^{\langle \dot \eta^\pm \rangle_{W_p,1} }
\nonumber \\ & = &
\langle \Delta \hat \rho^{(n)} \Delta \dot\eta_q^\pm  \rangle_{W_p,1} \,
e^{\langle \dot \eta^\pm \rangle_{W_p,1} }.
\end{eqnarray}
The final equality in Eq.~(\ref{Eq:rho-l}) is justified
by the passage from the second to the third equality here.
The average of the density fluctuation without loops vanishes,
$\langle \Delta \hat \rho^{(n)}({\bf Q}^n) \rangle_{W_p,1} = 0$.
Hence the only configurations that contribute
to the average in the second equality  in  Eq.~(\ref{Eq:exp-dot-eta})
are those with at least one loop in the vicinity of ${\bf Q}^n$.
This means that in any product of loops in the symmetrization function
$\eta^\pm_q({\Gamma})$,
one loop must be tied to ${\bf Q}^n$,
and the rest are free to wander throughout the volume.
Hence just as in the factorization of the grand partition function
in \S \ref{Sec:Omega},
the average of the density fluctuation times such a product
can be written as the product of the average of
the density fluctuation times the tied loop,
times the product of the averages of each of the free loops.
The symmetry number reflects the number and type of the free loops,
because these are equivalent and distinct from the tied loop.
Hence just as in \S \ref{Sec:Omega}, they sum to the exponential,
$ e^{ \langle \dot \eta_q^\pm \rangle_{W,1} }$,
which gives the third equality in  Eq.~(\ref{Eq:exp-dot-eta}).
The final equality in that equation,
combined with the prefactor
$\langle \eta_{q}^{\pm(l)}  \rangle_{W_p,1}^{-1}
= e^{- \langle \dot \eta_q^\pm \rangle_{W,1} }$
from the penultimate equality in Eq.~(\ref{Eq:rho-l}),
gives the final equality in Eq.~(\ref{Eq:rho-l}).

The result Eq.~(\ref{Eq:rho-l}) for the multi-particle position density
holds analogously for the multi-particle phase density,
Eq.~(\ref{Eq:rho-n-gamma}).
It follows that the average of any phase function
can be written in terms that involve
only the single loop symmetrization function,
as is now shown for the energy.

\subsubsection{Average Energy Factorized} \label{Sec:dens-H}

Suppose that the  Hamiltonian
consists of many-body potentials
${\cal H}({\bf \Gamma}) = \sum_n {\cal H}^{(n)}({\bf \Gamma})$,
with
\begin{eqnarray}
{\cal H}^{(n)}({\bf \Gamma})
& = &
\frac{1}{n!}
\sum_{k_1,\ldots,k_n}^N \!\!\!\!\!\!''\;\;
{\cal H}^{(n)}({\bf \Gamma}_{k_1},\ldots,{\bf \Gamma}_{k_n}).
\end{eqnarray}
The kinetic energy is included in the one-body term.

In terms of the position and momentum $n$-body density,
Eq.~(\ref{Eq:rho-n-gamma}),
the average Hamiltonian energy may be written
\begin{eqnarray}
\lefteqn{
\left\langle \hat {\cal H}  \right\rangle_{z,V,T}
} \nonumber \\
& = &
\sum_n
\frac{1}{n!} \int \mathrm{d}{\bm \gamma}^n  \;
\rho^{(n)}({\bm \gamma}^n)\, {\cal H}^{(n)}({\bm \gamma}^n)
\nonumber \\ & = &
\sum_n
\frac{1}{n!} \int \mathrm{d}{\bm \gamma}^n  \;
{\cal H}^{(n)}({\bm \gamma}^n)
\nonumber \\ &  & \mbox{ } \times
\left\{
\rho^{(n)}_{W_p,1}({\bm \gamma}^n)
+
\left\langle \Delta \rho^{(n)}({\bm \gamma}^n;{\bf \Gamma})
\Delta \dot\eta_q^\pm({\bf \Gamma})  \right\rangle_{W_p,1}
\right\}
\nonumber \\ & = &
\left\langle {\cal H}({\bf \Gamma}) \right\rangle_{W_p,1}
+
\left\langle \Delta {\cal H}({\bf \Gamma})
\Delta \dot\eta_q^\pm({\bf \Gamma}) \right\rangle_{W_p,1}
\nonumber \\ & = &
\left\langle {\cal H}({\bf \Gamma}) \right\rangle_{W_p,1}
+
\sum_{l=2}^\infty
\left\langle \Delta {\cal H}({\bf \Gamma})
\Delta \dot\eta_q^{\pm(l)}({\bf \Gamma}) \right\rangle_{W_p,1} .
\end{eqnarray}
This agrees with the expression obtained from
the temperature derivative of the grand potential,
\S \ref{Sec:Av-H-Omega},
where the loop contributions were given by
Eq.~(\ref{Eq:olEl-fluctn}),
\begin{equation}
\overline E_{W_p,l}
=
\left<  \left[ {\cal H}  - \overline E_{W_p,1} \right]
\left[ \dot\eta_{q}^{\pm(l)}
- \langle  \dot\eta_{q}^{\pm(l)} \rangle_{W_p,1}
\right]  \right>_{W_p,1} ,
\end{equation}
for $l \ge 2$.

%
\section{Virial Pressure}
\setcounter{equation}{0} \setcounter{subsubsection}{0}
%

The thermodynamic pressure is the volume derivative
of the grand potential\cite{Pathria72,TDSM}
\begin{equation}
\overline p^\pm
=
\frac{- \partial \Omega^\pm}{ \partial V}
=
\sum_{l =1}^\infty
\frac{- \partial \Omega^\pm_{l} }{ \partial V}.
\end{equation}

Perform the usual trick of scaling the position coordinates by
the edge length $L$, ${\bf q}' = {\bf q}/L$,
so that $\mathrm{d}{\bf q} =  {\bf q} L^{-1}\mathrm{d}L$.
Recall that the spacing between momentum states is
$\Delta_p = 2\pi \hbar /L$,
and so $\mathrm{d}{\bf p} = -{\bf p}L^{-1}  \mathrm{d}L$.
These give the change in a phase function as
\begin{eqnarray}
\mathrm{d}f({\bf \Gamma})
& = &
\nabla_p f \cdot \mathrm{d}{\bf p}
+ \nabla_q f \cdot  \mathrm{d}{\bf q}
\nonumber \\ & = &
\left\{ {\bf q} \cdot \nabla_q f - {\bf p} \cdot \nabla_p f  \right\}
L^{-1}  \mathrm{d}L.
\end{eqnarray}
In terms of the partition function,
the pressure is
\begin{eqnarray}
\beta \overline p^\pm
& = &
\frac{ \partial \ln \Xi^\pm}{ \partial V}
\nonumber \\ & = &
\frac{L}{dV \Xi^\pm}
\sum_{N=0}^\infty \frac{z^N}{h^{dN}N!}
\int \mathrm{d}{\bf \Gamma}\;
\nonumber \\ && \mbox{ } \times
\frac{\mathrm{d}}{\mathrm{d} L}
\left\{
e^{-\beta {\cal H}({\bf \Gamma})}
W_{p}({\bf \Gamma})  \eta_{q}^{\pm}({\bf \Gamma})
\right\}.
\end{eqnarray}

One has
\begin{eqnarray}
\frac{\mathrm{d} V^{N/2}
\langle\hat{\mathrm P}{\bf p}|{\bf q}\rangle}{\mathrm{d} L}
& = &
\frac{\mathrm{d} }{\mathrm{d} L}
e^{(\hat{\mathrm P}{\bf p})\cdot{\bf q}/i\hbar}
\nonumber \\ & = &
\frac{1}{i\hbar}
\left\{ {\bf q} \cdot (\hat{\mathrm P} {\bf p})
- (\hat{\mathrm P} {\bf p}) \cdot {\bf q} \right\}
 e^{(\hat{\mathrm P}{\bf p})\cdot{\bf q}/i\hbar}
 \nonumber \\ & = &
 0.
\end{eqnarray}
It follows that
$ {\mathrm{d} \eta_{q}^{\pm}({\bf \Gamma}) }/{\mathrm{d} L}
= 0 $,
and that therefore
\begin{eqnarray}
\lefteqn{
L \frac{\mathrm{d}}{\mathrm{d} L}
\left\{
e^{-\beta {\cal H}({\bf \Gamma})}
W_{p}({\bf \Gamma})
\eta_{q}^{\pm}({\bf \Gamma})
\right\}
}  \\
& = &
\eta_{q}^{\pm}({\bf \Gamma})
\left\{ {\bf q} \cdot \nabla_q  - {\bf p} \cdot \nabla_p   \right\}
\left\{
e^{-\beta {\cal H}({\bf \Gamma})} W_{p}({\bf \Gamma})
\right\}. \nonumber
\end{eqnarray}

The commutation function is given by
\begin{eqnarray}
e^{-\beta {\cal H}({\bf \Gamma})} W_{p}({\bf \Gamma})
& = &
\frac{ \langle {\bf q} | e^{-\beta \hat{\cal H}} | {\bf p} \rangle
}{ \langle {\bf q} | {\bf p} \rangle }
 \\ & = &
e^{ {\bf p} \cdot {\bf q} /i\hbar}
\int \mathrm{d}{\bf r}\;
\delta({\bf r}-{\bf q})
e^{-\beta \hat{\cal H}({\bf r})}
e^{ -{\bf p} \cdot {\bf r} /i\hbar } .\nonumber
\end{eqnarray}
The prefactor is a constant with respect to $L$,
and so one has
\begin{eqnarray}
\lefteqn{
L \frac{\mathrm{d}}{\mathrm{d} L}
e^{-\beta {\cal H}({\bf \Gamma})} W_{p}({\bf \Gamma})
} \nonumber \\
& = &
{\bf q} \cdot \nabla_q
\left\{
e^{-\beta {\cal H}({\bf \Gamma})} W_{p}({\bf \Gamma})
\right\}
- {\bf p} \cdot \nabla_p
\left\{
e^{-\beta {\cal H}({\bf \Gamma})} W_{p}({\bf \Gamma})
\right\}
\nonumber \\ & = &
e^{ {\bf p} \cdot {\bf q} /i\hbar}
\int \mathrm{d}{\bf r}\;
\left\{
{\bf q} \cdot \delta'({\bf r}-{\bf q})
e^{-\beta \hat{\cal H}({\bf r})}
e^{ -{\bf p} \cdot {\bf r} /i\hbar }
\right. \nonumber \\ && \left.  \mbox{ }
+ (i\hbar)^{-1}
\delta({\bf r}-{\bf q})
e^{-\beta \hat{\cal H}({\bf r})}
e^{ -{\bf p} \cdot {\bf r} /i\hbar }
{\bf p}\cdot {\bf r}
\right\}
\nonumber \\ & = &
-e^{ {\bf p} \cdot {\bf q} /i\hbar}
\int \mathrm{d}{\bf r}\;
\left\{\delta({\bf r}-{\bf q})
{\bf q} \cdot \nabla_{\bf r}
\left[
e^{-\beta \hat{\cal H}({\bf r})}
e^{ -{\bf p} \cdot {\bf r} /i\hbar } \right]
\right. \nonumber \\ && \left.  \mbox{ }
- (i\hbar)^{-1}
\delta({\bf r}-{\bf q})
e^{-\beta \hat{\cal H}({\bf r})}
e^{ -{\bf p} \cdot {\bf r} /i\hbar }
{\bf p}\cdot {\bf r}
\right\}
\nonumber \\ & = &
\frac{\langle {\bf q} | e^{-\beta \hat{\cal H}} {\bf p}\cdot{\bf r}
| {\bf p} \rangle
}{i\hbar  \langle {\bf q} | {\bf p} \rangle }
 -
 \frac{\langle {\bf q} |
 {\bf q}\cdot  \nabla_{\bf r} e^{-\beta \hat{\cal H}}
| {\bf p} \rangle
}{ \langle {\bf q} | {\bf p} \rangle }
\nonumber \\ & \equiv &
e^{-\beta{\cal H}({\bf \Gamma})}  W_p({\bf \Gamma})
{\cal V}_p({\bf \Gamma}).
\end{eqnarray}
The final equality defines what might be called
the quantum phase space virial.
Note that the ${\bf p}$ that appears explicitly
in the middle of the first term
to the right of the penultimate equality
is the momentum variable, not the momentum operator,
and that  the gradient operator $\nabla_{\bf r}$
that appears explicitly in the second term acts on everything to its right.

With this the pressure is given by
\begin{eqnarray}
dV \beta \overline p^\pm
& = &
\frac{1}{\Xi^\pm} 
\sum_{N=0}^\infty \frac{z^N}{h^{dN}N!}
\int \mathrm{d}{\bf \Gamma}\;
\nonumber \\ && \mbox{ } \times
e^{-\beta {\cal H}({\bf \Gamma})}  W_p({\bf \Gamma})
{\cal V}_p({\bf \Gamma})  \eta_{q}^{\pm}({\bf \Gamma}) .
\end{eqnarray}
or, equivalently,
\begin{eqnarray}
dV \beta \overline p^\pm
& = &
\left\langle {\cal V}_p({\bf \Gamma}) \right\rangle_{W_p,1}
+
\left\langle \Delta {\cal V}_p({\bf \Gamma})
\Delta \dot\eta_q^\pm({\bf \Gamma}) \right\rangle_{W_p,1}
\\ & = &
\left\langle {\cal V}_p({\bf \Gamma}) \right\rangle_{W_p,1}
+
\sum_{l=2}^\infty
\left\langle \Delta {\cal V}_p({\bf \Gamma})
\Delta \dot\eta_q^{\pm(l)}({\bf \Gamma}) \right\rangle_{W_p,1} .\nonumber
\end{eqnarray}

In practice possibly the simplest way to obtain
the quantum phase space virial
will turn out to be from the right hand side of the first equality,
\begin{eqnarray}
\lefteqn{
e^{-\beta{\cal H}({\bf \Gamma})}
W_p({\bf \Gamma})
{\cal V}_p({\bf \Gamma})
} \nonumber \\
& = &
{\bf q} \cdot \nabla_q
\left\{
e^{-\beta {\cal H}({\bf \Gamma})} W_{p}({\bf \Gamma})
\right\}
- {\bf p} \cdot \nabla_p
\left\{
e^{-\beta {\cal H}({\bf \Gamma})} W_{p}({\bf \Gamma})
\right\}
\nonumber \\ & = &
- \beta W_{p}({\bf \Gamma})
e^{-\beta {\cal H}({\bf \Gamma})}
\left\{ {\bf q} \cdot \nabla_q U({\bf q})
- {\bf p} \cdot \nabla_p {\cal K}({\bf p}) \right\}
\nonumber \\ && \mbox{ }
+
e^{-\beta {\cal H}({\bf \Gamma})}
\left\{ {\bf q} \cdot \nabla_q W_{p}({\bf \Gamma})
- {\bf p} \cdot \nabla_p W_{p}({\bf \Gamma}) \right\}.
\end{eqnarray}
This is useful when the `bare' commutation function $W_{p}$
can be otherwise obtained and differentiated.

In the classical limit of the present expression,
$W_p({\bf \Gamma}) = 1$,
and assuming only an internal potential,
this reduces to
\begin{eqnarray}
\lefteqn{
e^{-\beta{\cal H}({\bf \Gamma})} {\cal V}^\mathrm{cl}_p({\bf \Gamma})
} \nonumber \\
& = &
\left\{
{\bf q} \cdot \nabla_q e^{-\beta {\cal H}({\bf \Gamma})}
- {\bf p} \cdot \nabla_p e^{-\beta {\cal H}({\bf \Gamma})}
\right\} e^{-\beta {\cal H}({\bf \Gamma})}
\nonumber \\ & = &
\left\{
- \beta {\bf q} \cdot \nabla_q  U^\mathrm{int}({\bf q})
+ \frac{ \beta p^2 }{m}
\right\} e^{-\beta {\cal H}({\bf \Gamma})}
\nonumber \\ & = &
 \left\{ \beta {\cal V}^\mathrm{cl,int}({\bf q})
 + \frac{ \beta p^2 }{m}
\right\} e^{-\beta {\cal H}({\bf \Gamma})} .
\end{eqnarray}
Therefore,  in the classical limit
$W_p({\bf \Gamma}) = \eta^\pm_q = 1$,
the pressure is given by
\begin{eqnarray}
\beta \overline p^\mathrm{cl}
& = &
\frac{1}{dV \Xi^\mathrm{cl}}
\sum_{N=0}^\infty \frac{z^N}{h^{dN}N!}
\int \mathrm{d}{\bf \Gamma}\;
e^{-\beta{\cal H}({\bf \Gamma})} {\cal V}^\mathrm{cl}_p({\bf \Gamma})
\nonumber \\ & = &
\frac{1}{dV}
\left< \beta {\cal V}^\mathrm{cl,int}({\bf q})
+ \frac{ \beta p^2 }{m} \right>_{z,T}
\nonumber \\ & = &
\frac{1}{dV}
\left< dN \right>_{z,T}
+ \frac{\beta}{dV}
\left<  {\cal V}^\mathrm{cl,int}({\bf q}) \right>_{z,T} .
\end{eqnarray}
This agrees with the known classical expression for the virial pressure.
\cite{Pathria72,TDSM}



\section{Conclusion}

This paper has formulated quantum statistical mechanics
as an integral over classical phase space.
Two quantum phase functions appeared:
the commutation function,
which accounted for the non-commutativity
and lack of simultaneity of the position and momentum operators,
and the symmetrization function,
which accounted for the full symmetrization of the wave function
and the consequent statistics of bosons and fermions.

In general a specific commutation function was required
for each operator being averaged.
However in some circumstances this could be replaced by
the `bare'  commutation function from the grand partition function.

It was also shown that in the thermodynamic limit,
$V \rightarrow \infty$, $z,T=$ const.,
the symmetrization function involved in the partition function
and the statistical averages
could be factored and exponentially re-summed.
This is extremely convenient computationally and analytically.
The factorization and resummation was shown to be rather general,
since it held for multi-particle phase space densities, \S \ref{Sec:rho}.

Eight distinct expressions for the average energy  were discussed
that depended on three binary choices:
either $W_p\eta^\pm_q$ or $W_q\eta^\pm_p$,
either $W_{{\cal H}} $ or $ W $,
and either the trace form for a statistical average,
Eqs~(\ref{Eq:<H>,WH}) and (\ref{Eq:<H>,W}),
or else the derivative of the grand potential with factorization ansatz,
Eqs~(\ref{Eq:olE1}) and (\ref{Eq:olEl}).
The analysis in \S \ref{Sec:<H>} showed the consistency
of the first two binary choices,
and that in \S \ref{Sec:dens-H}
showed the consistency of the latter choice.




\begin{thebibliography}{99}


\bibitem{QSM}
P. Attard,
\emph{Quantum Statistical Mechanics:
Equilibrium and Non-Equilibrium Theory from First Principles},
(IOP Publishing, Bristol, 2015).



\bibitem{Attard17}
P. Attard,
arXiv:1702.00096 (2017).

\bibitem{Attard16}
P. Attard,
arXiv:1609.08178v3  (2016).

\bibitem{STD2}
P. Attard,
\emph{Entropy Beyond the Second Law.
Thermodynamics and Statistical Mechanics
for Equilibrium, Non-Equilibrium, Classical, and Quantum Systems},
(IOP Publishing, Bristol, 2018).


\bibitem{Bloch08}
I. Bloch, J. Dalibard, and W. Zwerger,
Rev.\ Mod.\ Phys.\ {\bf 80}, 885 (2008).


\bibitem{Hernando13}
A. Hernando and J. Van\'i\v cek,
Phys.\ Rev.\ A {\bf 88}, 062107 (2013).
arXiv:1304.8015v2 [quant-ph] (2013).


\bibitem{Georgescu11a}
I. Georgescu and V. A. Mandelshtam,
J. Chem.\ Phys.\ {\bf  135}, 154106 (2011).
arXiv:1107.3330v2 (2011).




\bibitem{Messiah61}
A. Messiah,
\emph{Quantum Mechanics},
(North-Holland, Amsterdam, Vols I and II, 1961).


\bibitem{Merzbacher70}
E. Merzbacher,
\emph{Quantum Mechanics},
(Wiley, New York, 2nd ed., 1970).

\bibitem{Pathria72}
R. K. Pathria,
\emph{Statistical Mechanics},
(Pergamon Press, Oxford, 1972).

\bibitem{Wigner32}
E. Wigner,
Phys.\ Rev.\ {\bf 40}, 749, (1932).

\bibitem{Kirkwood33}
J. G. Kirkwood,
Phys.\ Rev.\ {\bf 44}, 31, (1933).


\bibitem{Attard18}
P. Attard,
``Algorithm for Quantum Many-Particle Systems
by Transformation to Classical Phase Space
with Test Results for Quantum Harmonic Oscillators'', (2018).


\bibitem{TDSM}
P. Attard,
\emph{Thermodynamics and Statistical Mechanics:
Equilibrium by Entropy Maximisation}
(Academic Press, London, 2002).



\end{thebibliography}
\end{document}